\def\year{2018}\relax
\documentclass[letterpaper]{article} 
\usepackage{aaai18}  
\usepackage{times}  
\usepackage{helvet}  
\usepackage{courier}  
\usepackage{url}  
\usepackage{graphicx}  
\usepackage[fleqn]{amsmath}
\usepackage{microtype}
\usepackage{multirow}
\usepackage{subfig}
\usepackage{amssymb}
\usepackage{nccmath}
\usepackage{array}
\usepackage{comment}
\usepackage{color,soul}
\usepackage{upquote}
\usepackage{float}
\usepackage[hang,flushmargin]{footmisc}
\title{Cascade and Parallel Convolutional Recurrent Neural Networks on EEG-based Intention Recognition for Brain Computer Interface}
\author{Dalin Zhang,\textsuperscript{1}\footnotemark[1] Lina Yao,\textsuperscript{1} Xiang Zhang,\textsuperscript{1} Sen Wang,\textsuperscript{2} Weitong Chen,\textsuperscript{3} Robert Boots\textsuperscript{4, 5}\\
\textsuperscript{1}School of Computer Science and Engineering, The University of New South Wales, Australia\\
\textsuperscript{2}School of Information and Communication Technology, Griffith University, Australia\\
\textsuperscript{3}School of Information Technology and Electrical Engineering, The University of Queensland, Australia\\
\textsuperscript{4}Thoracic Medicine, Royal Brisbane and Women's Hospital, Australia\\
\textsuperscript{5}Burns Trauma and Critical Care Research Centre, The University of Queensland, Australia\\
\footnotemark[1]dalin.zhang@student.unsw.edu.au
}
\begin{document}
\maketitle

\begin{abstract}
Brain-Computer Interface (BCI) is a system empowering humans to communicate with or control the outside world with exclusively brain intentions. Electroencephalography (EEG) based BCIs are promising solutions due to their convenient and portable instruments. Despite the extensive research of EEG in recent years, it is still challenging to interpret EEG signals effectively due to the massive noises in EEG signals (e.g., low signal-noise ratio and incomplete EEG signals), and difficulties in capturing the inconspicuous relationships between EEG signals and certain brain activities. Most existing works either only consider EEG as chain-like sequences neglecting complex dependencies between adjacent signals or requiring preprocessing such as transforming EEG waves into images. In this paper, we introduce both cascade and parallel convolutional recurrent neural network models for precisely identifying human intended movements and instructions by effectively learning the compositional spatio-temporal representations of {\em raw} EEG streams. Extensive experiments on a large scale movement intention EEG dataset ({\bf 108} subjects, 3,145,160 EEG records) have demonstrated that both models achieve high accuracy near 98.3\% and outperform a set of baseline methods and most recent deep learning based EEG recognition models, yielding a significant accuracy increase of {\bf 18\%} in the {\em cross-subject} validation scenario. The developed models are further evaluated with a real-world BCI and achieve a recognition accuracy of 93\% over five instruction intentions. This suggests the proposed models are able to generalize over different kinds of intentions and BCI systems.
\end{abstract}

\section{Introduction}
\noindent Brain-computer interface (BCI) enables users to directly communicate with the outside world or to control instruments using brain intentions alone, thus providing an alternatively practical way to help people who are suffering from severe motor disabilities. Recent research has also found its applications for healthy users, such as BCI games in entertainment industries \cite{ahn2014review}. Scalp-recording electroencephalography (EEG) is considered to be one of the most practical pathways to realize BCI systems due to its portable acquisition system and convenient implementation \cite{wang2014asynchronous}. When a person {\em imagines} moving different parts of his body or different controlling commands for an instrument, the EEG signals from his scalp fluctuates in different modes. In this way, human intentions can be recognized by analyzing the EEG signals. It has been attracting increasing attentions, and various research has attempted to engage EEG based BCI in real-world applications such as mind controlled wheelchairs \cite{wang2014asynchronous}, prosthetic \cite{bright2016eeg} and exoskeletons \cite{qiu2017brain}.

\begin{figure*}[htb!]
\centering
\includegraphics[width=1\textwidth]{"arch"}
\caption{EEG data acquisition and preprocessing. EEG signals are first captured using a BCI headset with multiple electrodes and recorded as time series data vectors. These data vectors are then converted to 2D data meshes according to the electrode map of the BCI headset. The converted 2D meshes are finally segmented to clips using sliding window techniques.}
\label{fig:data precess}
\end{figure*}

However, real-world EEG based BCI systems are still immature due to diverse open challenges. First, EEG signals usually have a mass of noises.
Apart from the common noises of sensory systems, such as power line interference or inappropriate electrode connections, EEG signals have some unique inevitable noises. During the recording process, physiological activities like eye blinks, muscle activity and heart beat are all harm to collecting high signal-noise ratio EEG signals. It is hard to make sure that the participants concentrate on the performing tasks during the whole experiment period. Also, a typical EEG based BCI system usually has 8 to 128 signal channels resulting in limited signal resolution compared to image or video related tasks. Second, the correlations between the EEG signals and their corresponding brain intentions in deep structures are ambiguous. Unlike the body actions which can be easily explained by monitoring accelerometers or gyroscopes, it is not straightforward to infer the brain intentions by directly observing EEG signals. Third, widely utilized brain intention recognition methods heavily rely on handcrafted features, requiring extensive preprocessing before making a prediction \cite{sun2014review}. Some methods include signal de-noising \cite{heydari2015adaptive} or feature selection steps \cite{yin2017cross} followed by final recognition model. Such a two-stage model is inconvenient to train and implement, and the whole process is time-consuming and highly dependent on professional knowledge in this domain. Finally, current work mainly targets either intra-subject (test data and train data are from the same subject) or binary EEG signal classification scenarios. Little research has been carried out on both cross-subject and multi-class scenarios. However, the cross-subject and multi-class scenarios are highly desired for implementing real-world applications. Furthermore, even under intra-subject or binary classification scenarios, many existing works suffer poor performance near 80\% accuracy.

In recent years, deep learning's revolutionary advances in audio and visual signals recognition have gained significant attentions \cite{lecun2015deep}. Some recent deep learning based EEG classification approaches have enhanced the recognition accuracy \cite{pouya2016learning,tabar2016novel}. However, these approaches either focus on complex preprocessing, such as converting raw EEG signals to images \cite{pouya2016learning}, or neglecting the subtle spatial and temporal information contained within EEG signals. Hence, current methods still have limited capabilities in dealing with cross-subject and multi-class scenarios.

To tackle the above obstacles for further developing EEG-based BCIs, we present in this paper, two kinds of convolutional recurrent neural networks, which we call cascade and parallel models, 
to detect human intentions through learning the effective compositional spatio-temporal dynamics from {\em raw} EEG streaming signals without preprocessing. In particular, we build a mesh-like raw EEG signal hierarchy from 1D chain-like EEG vectors by mapping the EEG recordings with the spatial information of EEG acquisition electrodes, to align the correlations between neighbouring EEG signals and corresponding brain areas. Next, both cascade and parallel convolutional recurrent network models are developed to decode robust EEG representations from both space and time dimensions in sequence or in parallel respectively. The proposed models are unified end-to-end trainable models, simultaneously learning the robust feature representations and classifying the EEG raw signals to detect movement or instruction intentions. The proposed models have good generalization in more complex and practical scenarios (both cross-subject and multi-class). Both the cascade and parallel models achieve high accuracy of near 98.3\% for movement intention recognition, significantly outperforming the state-of-the-art methods by near 18\%. We also evaluate our models on a real-world BCI system, and obtain a satisfactory accuracy of 93\% on recognizing five instruction intentions with limited EEG channels. This reveals that our proposed models have robust capabilities to recognize diverse kinds of human intentions using different BCI systems.

\section{The Proposed Method}
In this section, we describe the detailed architectures of the proposed cascade and parallel convolutional recurrent network approaches. 
\subsection{Converting 1D EEG Sequences to 2D EEG Meshes}
The overall EEG data acquisition and preprocessing flowchart of our proposed method is shown in Figure \ref{fig:data precess}. The EEG based BCI system uses a wearable headset with multiple electrodes to capture the EEG signals. When a subject {\em imagines} performing a certain instruction, the electrodes of the headset acquire the fluctuations of the voltages from the scalp. The EEG electrode map in Figure \ref{fig:data precess} depicts the electrodes placement of an example BCI headset.
The electrode map varies from different BCI systems according to the different number of recording channels. The sensory readings from the EEG acquisition system represent time series data at the acquiring frequency. Typically, the raw data from EEG signal acquisition system at time index \textbf{\textit{t}} is a one-dimensional (1D) data vector $\textbf{r}_{t} = [s_{t}^{1},\enspace s_{t}^{2}\enspace, s_{t}^{i}...\enspace, s_{t}^{n}]^{T}$, where $s_{t}^{i}$ is the reading data of the $i$th electrode channel at time stamp $t$. The acquisition system totally contains $n$ channels. For the observation period $[t,\enspace t+N]$, there are $(N+1)$ 1D data vectors, each of which contains $n$ elements corresponding to $n$ electrodes of the acquisition headset. 

From the EEG electrode map, it is observed that each electrode is physically neighboring multiple electrodes which measures the EEG signals in a certain area of brain, while the elements of the chain-like 1D EEG data vectors are restricted to two neighbors. Furthermore, different brain regions correspond to different brain activities. From this conceptualization, we convert the 1D EEG data vectors to 2D EEG data meshes according to the spatial information of the electrode distribution of the acquisition system. The transformation function of the 1D data vector $\textbf{r}_{t}$ at time stamp $t$ for its corresponding 2D data mesh $\textbf{m}_{t}$ is denoted as follows: 

\begin{equation}\label{eq:1}
\textsl{T}(\textbf{r}_{t}) = \left[ \begin{smallmatrix} 0&0&0&0&s_{t}^{22}&s_{t}^{23}&s_{t}^{24}&0&0&0&0\\  0&0&0&s_{t}^{25}&s_{t}^{26}&s_{t}^{27}&s_{t}^{28}&s_{t}^{28}&0&0&0\\
0&s_{t}^{30}&s_{t}^{31}&s_{t}^{32}&s_{t}^{33}&s_{t}^{34}&s_{t}^{35}&s_{t}^{36}&s_{t}^{37}&s_{t}^{38}&0\\
0&s_{t}^{39}&s_{t}^{1}&s_{t}^{2}&s_{t}^{3}&s_{t}^{4}&s_{t}^{5}&s_{t}^{6}&s_{t}^{7}&s_{t}^{40}&0\\
s_{t}^{43}&s_{t}^{41}&s_{t}^{8}&s_{t}^{9}&s_{t}^{10}&s_{t}^{11}&s_{t}^{12}&s_{t}^{13}&s_{t}^{14}&s_{t}^{42}&s_{t}^{44}\\
0&s_{t}^{45}&s_{t}^{15}&s_{t}^{16}&s_{t}^{17}&s_{t}^{18}&s_{t}^{19}&s_{t}^{20}&s_{t}^{21}&s_{t}^{46}&0\\
0&s_{t}^{47}&s_{t}^{48}&s_{t}^{49}&s_{t}^{50}&s_{t}^{51}&s_{t}^{52}&s_{t}^{53}&s_{t}^{54}&s_{t}^{55}&0\\
0&0&0&s_{t}^{56}&s_{t}^{57}&s_{t}^{58}&s_{t}^{59}&s_{t}^{60}&0&0&0\\
0&0&0&0&s_{t}^{61}&s_{t}^{62}&s_{t}^{63}&0&0&0&0\\
0&0&0&0&0&s_{t}^{64}&0&0&0&0&0\\ \end{smallmatrix} \right]
\end{equation}

\noindent where the positions of the \textit{null} electrodes are padding with zeros. Through this transformation, the raw 1D data vector series $[\textbf{r}_{t},\enspace\textbf{r}_{t+1}\enspace...\enspace\textbf{r}_{t+N}]$ is converted to the 2D data mesh series $[\textbf{m}_{t},\enspace\textbf{m}_{t+1}\enspace...\enspace\textbf{m}_{t+N}]$. During observation duration $[t,\enspace t+N]$, the number of 2D data meshes is still $(N+1)$. After 2D data mesh transformation, the data mesh is normalized across the non-zero elements using Z-score normalization. Each of the resulted 2D data meshes contains the spatial information of the brain activity at its recording time. During the recording process, some EEG readings are variably missing largely due to issues of electrical conductivity and subjects movement, resulting in all channels recording zeros. This issue is unavoidable in sensor-based systems, and it might not be tolerated by BCIs. From the application point of view, smooth manipulation of the BCI system provides improved user experience. For this reason, a BCI system should preferably translate brain activities to the output information continuously without interruption. As missing information is a clinical reality, in this work we preserve the incomplete recordings which are discarded in previous work \cite{kim2016motor} to maintain the integrity of EEG signals. The experimental results show our 2D EEG meshes perform well in dealing with the ``missing readings".

Up to this point, we apply the sliding window approach to divide the streaming 2D meshes to individual clips as shown in the last step of Figure \ref{fig:data precess}. Each clip has fixed length of time series 2D data meshes with 50\% overlapping between continuous neighbors. The data meshes segment $\textbf{S}_j$ is created as follows:

\begin{ceqn}
\begin{align*}
\textbf{S}_{j} = [\textbf{m}_{t}, \enspace\textbf{m}_{t+1}\enspace...\enspace\textbf{m}_{t+S-1}]
\end{align*}
\end{ceqn}
where $S$ is the window size and $j = 1,2,...,q$ with $q$ segments during the observation period. Our goal is to develop an effective model to recognize a set of human intentions $\textbf{A} = [a_{1},\enspace a_{2}\enspace...\enspace a_{K}]^{T}$ from each windowed data meshes segment $\textbf{S}_{j}$. The recognition approach tries to predict the human intention $\textbf{Y}_{t}\in\textbf{A}$ performed during this windowed period.

\begin{figure}[htb]
\centering
\includegraphics[width=.47\textwidth]{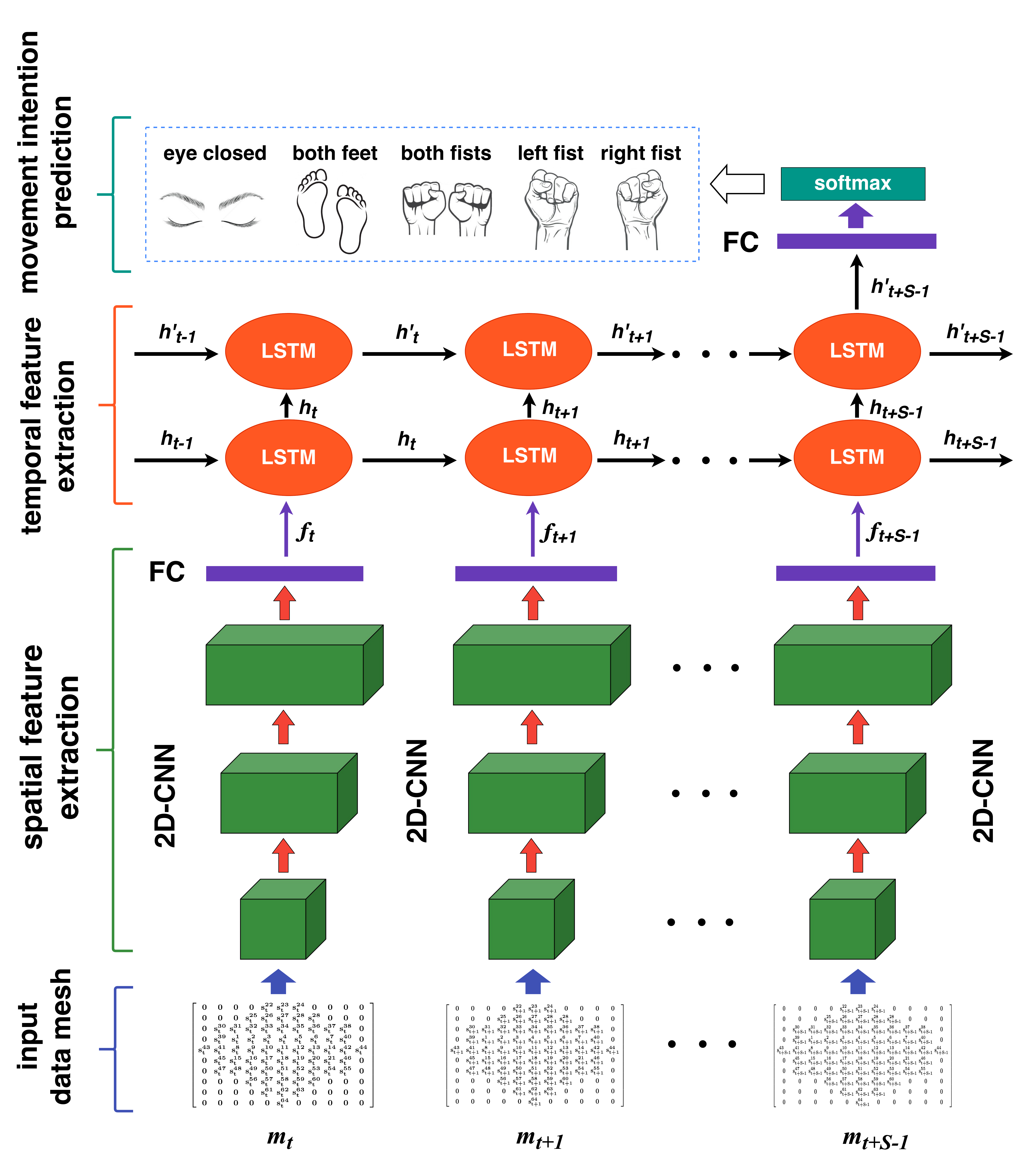}
\caption{Cascade convolutional recurrent neural network architecture.}
\label{fig:cascade}
\end{figure}

\subsection{Cascade Convolutional Recurrent Network}
We first design a cascade deep convolutional recurrent neural network framework illustrated in Figure \ref{fig:cascade}, capturing the spatial and temporal features in EEG sequences. The input to the model is the preprocessed segment of 2D data meshes (e.g., $\textbf{S}_{j}$), creating a 3D data architecture containing both spatial and temporal information. We first extract the spatial features of each data mesh, and then feed the sequence of the extracted spatial features into the RNN to extract temporal features. One fully connected layer receives the output of the last time step of the RNN layers, and feeds the softmax layer for final intention prediction.

To extract the spatial features of each data mesh, we apply a mesh-wise deep 2D-CNN as shown in Figure \ref{fig:cascade}. The $j$th input segment is defined as $\textbf{S}_{j}= [\textbf{m}_{t},\enspace\textbf{m}_{t+1}\enspace...\enspace\textbf{m}_{t+S-1}]\in \mathbb{R}^{S\times h \times w}$, where there are $S$ data meshes denoted as $\textbf{m}_k\enspace(k=t,\enspace t+1\enspace...\enspace t+S-1$), and each data mesh is of size $h\times w$. The data meshes are input to a 2D-CNN individually, and each resolves to a spatial feature representation $\textbf{f}_k\enspace(k=t,\enspace t+1\enspace...\enspace t+S-1)$:

\begin{ceqn}
\begin{align*}
    \text{CasCNN:}\enspace\textbf{f}_k &= C_{2D}(\textbf{m}_k),\enspace\textbf{f}_k\in \mathbb{R}^{l}.
\end{align*}
\end{ceqn}

\noindent The final spatial feature representation $\textbf{f}_k$ is a feature vector with $l$ elements. Through the 2D-CNN spatial feature extraction step, the input segments are transformed to sequences of spatial feature representations:
\begin{ceqn}
\begin{align*}
    \text{CasCNN:}\enspace\textbf{S}_{j} \Rightarrow \textbf{F}_{j},
    \text{where}\enspace\textbf{F}_{j}= [\textbf{f}_{t}\enspace...\enspace\textbf{f}_{t+S-1}]\in\mathbb{R}^{S\times l}.
\end{align*}
\end{ceqn}

Concretely, there are three 2D convolutional layers with the same kernel size of $3 \times 3$ for spatial feature extraction. In each convolutional operation we use zero-padding techniques to prevent missing the information at the edge of the input data mesh. This creates feature maps with the same size as the raw input EEG data mesh of $h\times w$. We start the first convolutional layer with 32 feature maps, and double the feature maps in each of the following convolutional layers. As a result, there are 128 feature maps in the last convolutional layer. After these three convolutional layers, a fully connected layer with 1024 neurons is applied to convert the 128 feature maps to the final spatial feature representation $\textbf{f}_k\in \mathbb{R}^{1024}$. This fully connected layer is optional for feeding the 2D-CNN results to RNN. However, we observe that this layer is essential in helping with convergence and marginally improvement of the performance of the whole framework. 

The spatial feature representation sequence $\textbf{F}_j$ is input to a RNN to computes the temporal features. We use Long Short-Term Memory (LSTM) units to construct two stacked RNN layers. LSTM is a modified RNN cell addressing the gradient vanishing and exploding problem. There are $S$ LSTM units in each layer, and the input to the second RNN layer is the output time sequence of the previous RNN layer. The hidden state of the LSTM unit of the first RNN layer at current time step $t$ is denoted as $h_t$, and the $h_{t-1}$ is the hidden state of the previous time step $t-1$. The information from the previous time step is conveyed to the current step, and influence the final output. We use the hidden state of the LSTM unit as the output of the LSTM unit. Therefore, the input sequence of the second LSTM layer, is the hidden state sequence of the first LSTM layer $[\textbf{h}_t,\enspace \textbf{h}_{t+1}\enspace...\enspace \textbf{h}_{t+S-1}]$. Since we are interested in what the brain is directing during the whole segment period, the extracted features when the LSTM has observed the entire samples of the sliding window are used for further analysis. Only the output of the last time step LSTM, $\textbf{h}'_{t_S-1}$, is fed into the next fully connected layer as shown in the final stage of Figure \ref{fig:cascade}. The temporal feature representation $\textbf{h}'_{t+S-1}$ of the segment $\textbf{S}_j$ is:

\begin{ceqn}
\begin{align*}
    \text{CasRNN:}\enspace\textbf{h}'_{t+S-1} &= R_{lstm}(\textbf{F}_j),\enspace\textbf{h}'_{t+S-1}\in \mathbb{R}^{d},
\end{align*}
\end{ceqn}
where $d$ is the size of the hidden state of an LSTM unit. On top of the fully connected layer is the final softmax layer yielding final probability prediction of each class:
\begin{ceqn}
\begin{align*}
    \text{FC-softmax:}\enspace\textbf{P}_{j} &= S_m(\textbf{h}'_{t+S-1}),\enspace\textbf{P}_{j}\in \mathbb{R}^{K},
\end{align*}
\end{ceqn}
where the framework aims to classify $K$ categories. We induce dropout operations as a form of regularization after the fully connected layers in both the 2D-CNN stage and the final classification stage. 

Overall, the framework converts and splits the EEG recording streams to segments of 2D data meshes, and classifies each segment to one of the $K$ categories. Each segment $\textbf{S}_j$ contains $S$ EEG data recordings, which have been converted to $S$ 2D meshes $[\textbf{m}_{t},\enspace\textbf{m}_{t+1}\enspace...\enspace\textbf{m}_{t+S-1}]$. A 2D-CNN is applied mesh-wise in a segment to extract spatial features $[\textbf{f}_{t},\enspace\textbf{f}_{t+1}\enspace...\enspace\textbf{f}_{t+S-1}]$, and a RNN is consequently applied to extract the temporal features $\textbf{h}'_{t+S-1}$ across the data meshes. Softmax classifier finally computes the classification probabilities over $K$ brain intentions for each individual segment. 

\begin{figure}[ht]
\centering
\includegraphics[width=.47\textwidth]{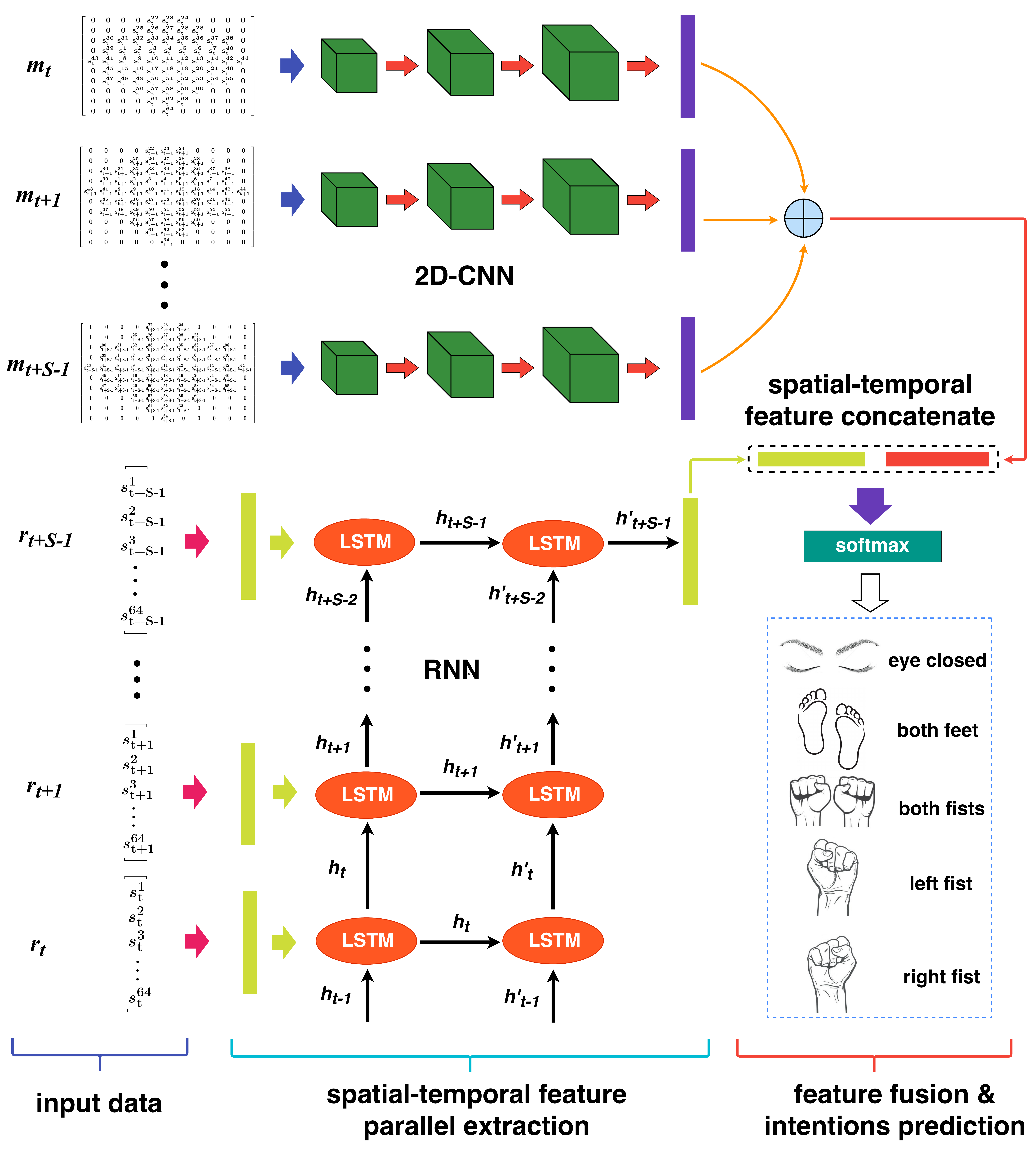}
\caption{Parallel recurrent convolutional neural network architecture. The concatenation operation in the final spatio-temporal fusion part is used for an example.}
\label{fig:parallel}
\end{figure}

\subsection{Parallel Convolutional Recurrent Network}
We also propose a parallel convolutional recurrent network. It is illustrated in Figure \ref{fig:parallel}. It also contains two parts, CNN and RNN, for spatial and temporal feature extraction respectively. However, different from the cascade model, the parallel model extracts the spatial and temporal features of EEG signals in parallel and fuses the extracted features at last for final intention recognition. Particularly, the RNN part of the parallel model receives the data from the same segments to that feed the corresponding CNN part. The $j$th input windowed segment to the RNN part is:
\begin{ceqn}
\begin{align*}
    \textbf{R}_{j} &= [\textbf{r}_{t}, \enspace\textbf{r}_{t+1}\enspace...\enspace\textbf{r}_{t+S-1}],
\end{align*}
\end{ceqn}
\noindent where $\textbf{r}_t$ is the data vector at time step $t$, and $S$ denotes the window size. The RNN part of the parallel model also has two LSTM layers, each containing the same number of LSTM units with that of the cascade model due to the same window size we use. The hidden state of the last time step in one segment is used for further analysis as well:
\begin{ceqn}
\begin{align*}
   \textbf{h}'_{t+S-1} &= R_{lstm}(\textbf{R}_j),\enspace\textbf{h}'_{t+S-1}\in \mathbb{R}^{v},
\end{align*}
\end{ceqn}
where $v$ is the hidden state size of the LSTM unit. A fully connected layer is applied both before and after LSTM layers to enhance the temporal information representation capabilities. Thus the final temporal features from the parallel RNN part is denoted as:
\begin{ceqn}
\begin{align*}
   \text{ParaRNN:}\textbf{O}_j &= \text{FC}(\textbf{h}'_{t+S-1}),\enspace\textbf{O}_j\in \mathbb{R}^{l},
\end{align*}
\end{ceqn}
where $l$ is the size of the finally fully connected layer of the parallel RNN part. The parallel CNN part which is responsible for extracting spatial features, receives the segment of 2D data meshes $\textbf{S}_{j}$ as input, and applies mesh-wise convolutional operations as the CNN part of the cascade model does. The CNN structure of the parallel model is the same with that of the cascade model as well. To be comparable to the temporal features in terms of size, the extracted spatial features $\textbf{f}_k$ at each time step in one segment are added up to a single feature vector $\textbf{L}_j$:
\begin{ceqn}
\begin{align*}
    \text{ParaCNN:}\enspace\textbf{L}_{j} &= \sum_{k=t}^{t+S-1}\textbf{f}_k\enspace (\textbf{L}_j,\enspace\textbf{f}_k\in\mathbb{R}^{l}),
\end{align*}
\end{ceqn}
\noindent where $l$ is the size of the fully connected layer in the CNN part, which is the same with that of the RNN part.

The concurrently extracted spatial and temporal features are fused to a joint spatio-temporal feature vector. Various fusion approaches are developed, and the detailed results are shown in the following experiment section. A softmax layer receives the fused spatio-temporal features to finally predict the human intentions:
\begin{ceqn}
\begin{align*}
    \text{softmax:}\enspace\textbf{P}_{j} &= S_m([\textbf{L}_j,\enspace\textbf{O}_j]),\enspace\textbf{P}_{j}\in \mathbb{R}^{K}.
\end{align*}
\end{ceqn}

In the 2D-CNN part of both the cascade and parallel models, convolutional layers are not followed by a pooling operation. Although in a typical CNN architecture a convolutional operation is often coupled with a pooling operation, it is not mandated. The pooling operation is usually used for reducing data dimensions at the cost of missing some information. However, in this EEG data analysis problem, the data dimension is much smaller than that used in computer vision research, so in order to keep all information, we concatenate three CNN layers directly without pooling operations. 

\section{Experiments and Result Summary}
We focus on the PhysioNet EEG Dataset \cite{goldberger2000physiobank} of the cross-subject, multi-class scenario to evaluate the proposed cascade and parallel models for movement intention recognition. The developed models are compared against those previous reported to show the superior performance. Meanwhile, we also systematically investigate the influence of the spatial and temporal information, and the performance of different variants of both cascade and parallel models. At last a case study using a real-world BCI system is conducted to evaluate the proposed models.

\subsection{Dataset and Model Implementation}
The movement intention EEG data is collected using BCI2000 instrumentation \cite{schalk2004bci2000} with 64 electrode channels and 160Hz sampling rate. To the best of our knowledge, this dataset is so far the largest EEG-based movement intention dataset with 109 subjects. But in the data preprocessing stage, we found that the labels of the \#89 subject were severely damaged, so this participant's record was removed from further analysis. We used the EEG data from 108 subjects to build the cross-subject dataset. The dataset contains five brain activities with eye closed (baseline), imagining moving both feet, both fists, left fist and right fist. 

All neural networks were implemented with the TensorFlow framework and trained on a Nvidia Titan X pascal GPU from scratch in a fully-supervised manner. The stochastic gradient descent with Adam update rule \cite{kingma2014adam}
is used to minimize the cross-entropy loss function. The network parameters are optimized with a learning rate of $10^{-4}$. The keep probability of the dropout operation is 0.5. According to the EEG data recording process of the evaluation dataset, the 2D data meshes are transformed with the size of $10\times11$ as shown in Figure \ref{fig:data precess}. The length of the window $S$ is set to 10. The hidden states of the LSTM cell for cascade model $d$ and parallel model $v$ are 64 and 16 respectively. All fully connected layers have the same size of 1024.

\subsection{Comparison Models}
\subsubsection{State-of-the-arts}
We will give a brief introduction of the compared state-of-the-art models. All the models are based on the same dataset with our work.

\begin{itemize}
\item \cite{major2017effects} researches independent component analysis (ICA) to reduce noises and feed the result to a neural network for final prediction on intra-subject binary MI-EEG classification;
\item \cite{shenoy2015shrinkage} uses shrinkage regularized filter bank common spatial patterns (SR-FBCSP) for intra-subject binary MI-EEG classification;
\item \cite{pinheiro2016wheelchair} focuses on one-against-all EEG classification using SVM, nearest neighbour and C4.5 algorithms;
\item \cite{kim2016motor} extracts EEG features with strong uncorrelating transform complex common spatial patterns (SUTCCSP) algorithm, and make final predictions with random forest classifier for the cross-subject binary classification;
\item \cite{xiang2017mobiquitous} uses autoencoder for EEG feature extraction and XGboost for final classification on five-category, cross-subject motor imagery scenario;
\item \cite{pouya2016learning} extracts the frequency features of EEG data, and converts the extracted features to images to feed into recurrent-convolutional
network. We reproduce their method on the same MI-EEG dataset with this work using their open access code. 
\end{itemize}

\subsubsection{Baseline models}
Apart from a set of state-of-the-arts, we also compare our model with the variants of CNN- and RNN-based models. We use 1D-CNN (without spatial or temporal information), 2D-CNN (only with spatial information) and 3D-CNN (with both spatial and temporal information) models for comparison and investigating the influence of spatial and temporal information on brain intention recognition. The 1D-CNN model just uses the raw EEG vectors as input. The 2D-CNN model is fed with the data meshes transformed by equation (\ref{eq:1}), but without sliding window segmentation. The 3D-CNN model uses the same input data with that fed into the cascade model. Each of the three CNN models has three convolutional layers without subsampling layers, one fully connected layer with 1024 neurons and one softmax output layer. The kernel size of the models are $3$, $3\times 3$ and $3\times 3\times 3$ for 1D, 2D and 3D, respectively, and the stride keeps constant of $1$. There are $32$, $64$ and $128$ feature maps in CNN for all baseline models. For comparison purpose, we keep all the hyper-parameters of the baseline CNN models the same with the CNN part of our proposed method. To make a fair comparison with both the cascade model and the parallel model, we use two RNN baseline models both with two LSTM layers between two fully connected layers, and choose 64 and 16 as the hidden state size of LSTM cells respectively.

\subsection{Experimental Results}
In this section, we will present the overall performance of our proposed models and the comparison results. The influence of spatial and temporal information, and variants of the proposed models will also be systematically analyzed.
\subsubsection{Overall Performance}
\begin{table}[!htb]
\centering
\begin{scriptsize}
    \caption{Comparison with the state-of-the-art methods and baseline methods. All the methods are based on the same dataset. RNN(64) and RNN(16) denote RNN models with hidden state size of 64 and 16 respectively. Cross-Sub (108) refers to the number of subjects included in the experiments.}
	\begin{tabular}{ >{\centering}m{2.8cm} ccc}
	    \hline
		\textbf{Method} &\textbf{Multi-class}&\textbf{Validation}& \textbf{Accuracy}\\
		\hline 
		\cite{major2017effects}&Binary&Intra-Sub&0.72\\
		\cite{shenoy2015shrinkage}&Binary&Intra-Sub&0.8206\\
        \cite{pinheiro2016wheelchair}&Binary&Cross-Sub(10)&0.8505\\
        \cite{kim2016motor}&Binary&Cross-Sub(105)&0.805\\
        \cite{xiang2017mobiquitous}&Multi(5)&Cross-Sub(20)&0.794\\
        \cite{pouya2016learning}\footnotemark&Multi(5)&Cross-Sub(108)&0.6731\\
        \hline
        1D-CNN&Multi(5)&Cross-Sub(108)&0.8622\\
        2D-CNN&Multi(5)&Cross-Sub(108)&0.8841\\
        3D-CNN&Multi(5)&Cross-Sub(108)&0.9238\\
        \hline
        RNN(64)&Multi(5)&Cross-Sub(108)&0.8493\\
        RNN(16)&Multi(5)&Cross-Sub(108)&0.7468\\
        \hline
        \textbf{Cascade model}&Multi(5)&Cross-Sub(108)&\textbf{0.9831}\\
        \textbf{Parallel model}&Multi(5)&Cross-Sub(108)&\textbf{0.9828}\\
        \hline
	\end{tabular}
    \label{tab: stat of art}
\end{scriptsize}
\end{table}

\footnotetext{We reproduce the approach on our dataset using the open access code on github https://github.com/pbashivan/EEGLearn}
The overall performance of our proposed models and the comparison models are summarized in Table \ref{tab: stat of art}. It is observed that both our cascade and parallel models achieve high accuracy near 98.3\%, consistently outperforming the state-of-the-art methods and the baseline models. Even though some work is focused on simple scenarios, such as intra-subject or binary classification, our method surpasses their methods significantly. Furthermore, our 3D-CNN baseline model also achieves competitive results to the state-of-the-art work. Bashivan et al. also proposed to use CNN and RNN for EEG signal analysis \cite{pouya2016learning}. However, they used complex preprocessing steps extracting the spectral features of EEG signals and converting to 2D images instead of using the raw signal data. To make a comparison, we reproduced their method on our dataset using their open access code on Github, and the results are also shown in Table \ref{tab: stat of art}. Our approach outperforms Bashavan's models by some 30\%. This is probably because their spectral feature extraction steps include a data compression process over a large continuous sampling period, while the movement intention tasks are periodic short duration brain activities. So extracting the spectral features may lose critical informative messages within the raw signals. What's more, they also use the interpolation approach to extend the raw 64-channel data to a $32\times32$ matrix, which brings in lots of noises. Compared with previous studies, our models directly utilize the raw EEG data, with no need for domain knowledge to select related frequency bands or complicated preprocess steps at the risk of missing critical information or introducing a mass of noises.
In addition, less preprocess making it more favourable for real-time applications, such as BCI.

\subsubsection{Impact of Temporal and Spatial Information}
To investigate the influence of spatial and temporal information on movement intention recognition, we build up CNN and RNN baseline models as as depicted above, and their performance is also summarized in Table \ref{tab: stat of art}. It is noted that increasing the CNN model's dimension, which means adding spatial or temporal information, obviously enhances the model's performance. What's more, sole CNN or RNN models are not able to achieve comparative performance with both the cascade and parallel models. It is also observed that the 3D-CNN model, which just represents the local temporal information, is not as powerful as the proposed models that involve the global temporal information by the RNN parts. These comparison results imply that it is crucial to use both the spatial and temporal information to boost EEG-based intention recognition and analysis. 

\subsubsection{Variants of Cascade and Parallel models}
\begin{table}
\centering
\begin{scriptsize}
    \caption{Comparison of different variants of cascade convolutional recurrent network model}
	\begin{tabular}{c|cc}
	    \hline
		\textbf{Cascade structure}& \textbf{Accuracy}& \textbf{F1 score}\\
		\hline
		1-layer CNN+FC+2-layer RNN+FC&0.9310&0.9207\\
		2-layer CNN+FC+2-layer RNN+FC&0.9712&0.9676\\
		\textbf{3-layer CNN+FC+2-layer RNN+FC}&\textbf{0.9831}&\textbf{0.9804}\\
		\hline
		3-layer CNN+2-layer RNN+FC&0.9217&0.9117\\
		3-layer CNN+FC+2-layer RNN&0.9801&0.9783\\
		\hline
		3-layer CNN+FC+1-layer RNN+FC&0.9813&0.9791\\
        \hline
	\end{tabular}
    \label{tab: cascade}
\end{scriptsize}
\end{table}

\begin{table}[!htb]
\centering
\begin{scriptsize}
    \caption{Comparison of different structures of parallel convolutional recurrent network model. Conv is short for point-wise convolutional operation.}
	\begin{tabular}{cc|cc}
	    \hline
		\textbf{Parallel structure}& \textbf{Fusion method}& \textbf{Accuracy}&  \textbf{F1 score}\\
		\hline
		
		1-layer CNN+FC&\multirow{2}{*}{Concatenation}&\multirow{2}{*}{0.9487}&\multirow{2}{*}{0.9432}\\
		FC+2-layer RNN+FC&&\\

		2-layer CNN+FC&\multirow{2}{*}{Concatenation}&\multirow{2}{*}{0.9727}&\multirow{2}{*}{0.9697}\\
		FC+2-layer RNN+FC&&\\

		\textbf{3-layer CNN+FC}&\multirow{2}{*}{\textbf{Concatenation}}&\multirow{2}{*}{\textbf{0.9828}}&\multirow{2}{*}{\textbf{0.9810}}\\
		\textbf{FC+2-layer RNN+FC}&&\\
		
		3-layer CNN+FC&\multirow{2}{*}{Concatenation}&\multirow{2}{*}{0.9821}&\multirow{2}{*}{0.9793}\\
		FC+1-layer RNN+FC&&\\
		\hline
		3-layer CNN+FC&\multirow{2}{*}{Summation}&\multirow{2}{*}{0.9813}&\multirow{2}{*}{0.9792}\\
		FC+2-layer RNN+FC&&\\
		\hline
		3-layer CNN+FC&\multirow{2}{*}{Concatenation+FC}&\multirow{2}{*}{0.9696}&\multirow{2}{*}{0.9661}\\
		FC+2-layer RNN+FC&&\\
		\hline
		3-layer CNN+FC&\multirow{2}{*}{Concatenation+Conv}&\multirow{2}{*}{0.9666}&\multirow{2}{*}{0.9626}\\
		FC+2-layer RNN+FC&&\\
		\hline
	\end{tabular}
    \label{tab: parallel}
\end{scriptsize}
\end{table}
Since it is impossible to exhaustively investigate the neural network architectures, here we study the effects of the key components of the proposed models. The results are summarized in Table \ref{tab: cascade} and Table \ref{tab: parallel} for the cascade model and the parallel model respectively. It is shown that more CNN or RNN layers would result better accuracy for both frameworks. However this performance improvement is at the cost of computational resources, thus we choose three CNN layers and two RNN layers by the trade-off between performance and efficiency. Fully connected layers are also critical components for the cascade model to create robust spatio-temporal representations, especially the layer linking the CNN part and the RNN part. In the parallel model, the data flows through the CNN and RNN concurrently, and there are diverse methods to fuse the parallel features. Here two basic fusion approaches (concatenation and summation) as well as two improved fusion approaches (concatenation joint fully connected layer and concatenation joint pointwise convolutional operation \cite{Chollet_2017_CVPR}) are studied. It is interesting to find that the basic fusion methods perform better results with accuracy higher than 98\%. Complex or advanced neural network needs careful training and parameter tuning to achieve better performance, thus it is redundant to add more operations when basic approaches are capable to achieve satisfactory results. 

\begin{figure}[htb]
\subfloat{
    \includegraphics[width=.195\textwidth]{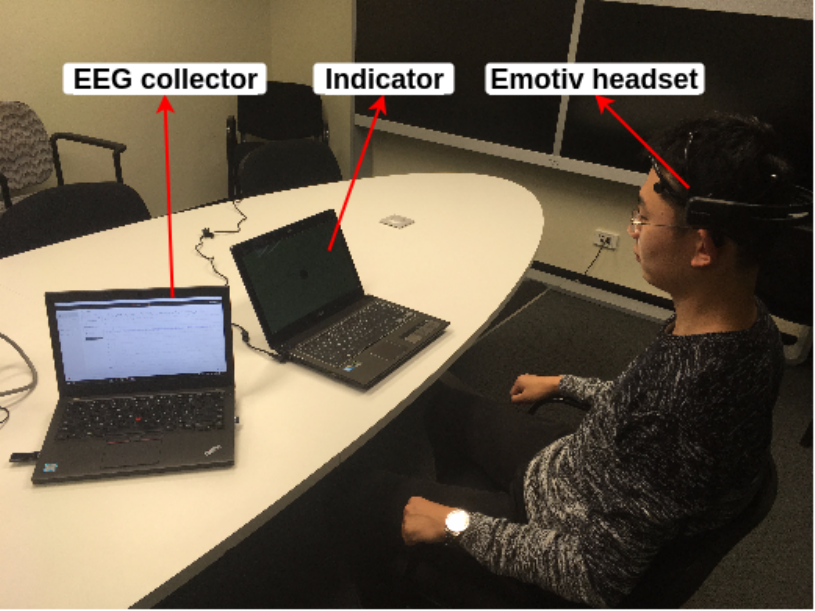}}
\subfloat{
    \includegraphics[width=.265\textwidth]{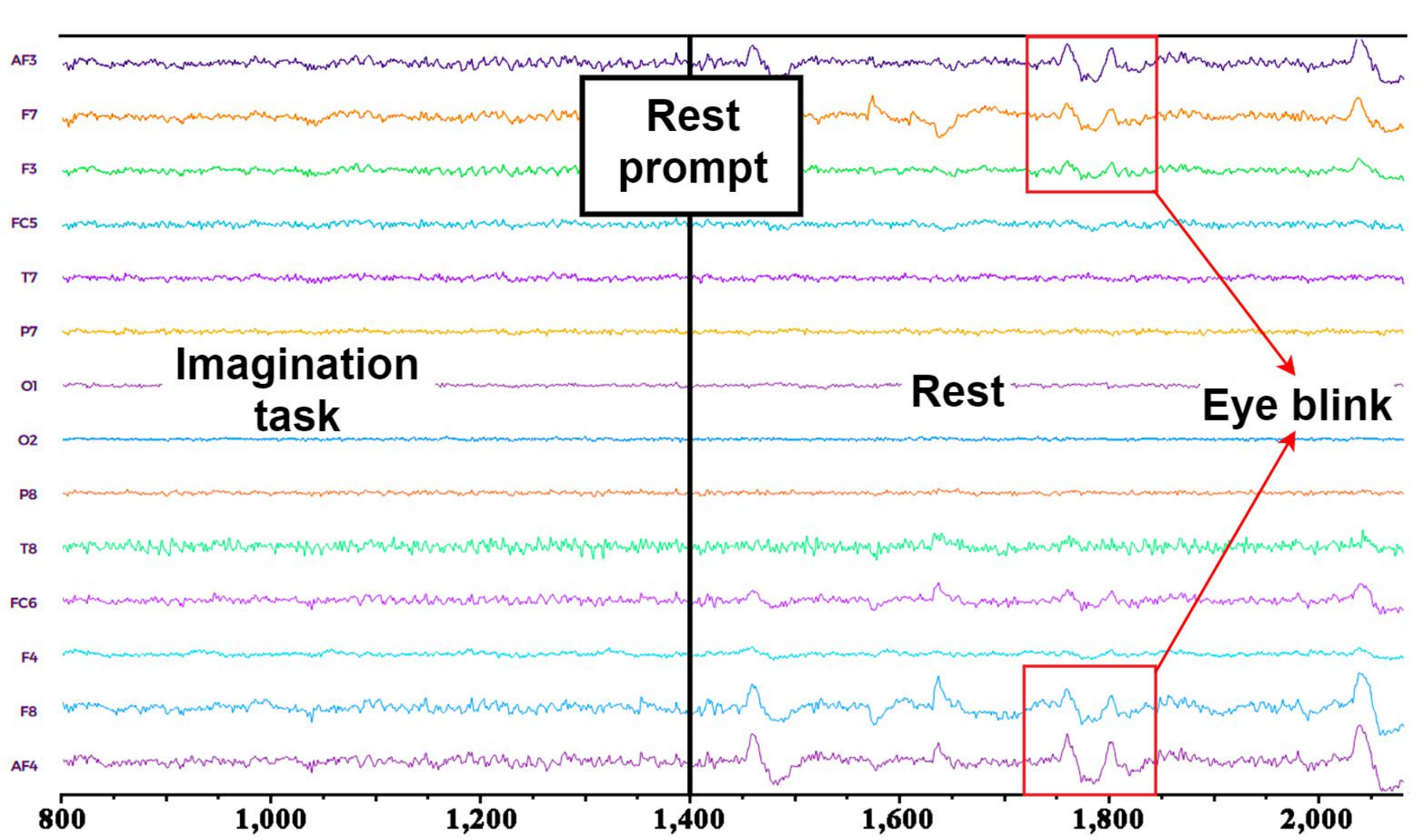}}
  \caption{EEG signal recording process (a) A participant performing the prompted intention task (b) Raw EEG signal recording interface}
  \label{fig:recording}
\end{figure}

\subsection{Case Study}
We evaluate the proposed models on our experimental dataset for instruction intention recognition. The 14-channel wireless EMOTIV Epoc+ EEG acquisition system was used to record raw EEG signals with sampling rate of 128Hz. The recording process is shown in Figure \ref{fig:recording}. Each participant executed five kinds of instruction intentions according to the prompts on the indicator in front of him. Arrows of four directions prompt the participants to perform intending to move the arrows to the corresponding directions, namely forward, backward, left and right. A circle prompts the participant to think nothing but to stare at the screen, representing \textit{null} intentions. In one recording trial, the participants perform an intention task for 10 seconds followed by a 10-second rest. Every volunteer performs 30 trials, and there are totally 9 volunteers including 3 females and 6 males. Finally we got 270 trials, 54 trials for each intention. All the recordings are mixed up to form a cross-subject multi-class dataset for further evaluating. During the experimental process, it is found that physiological activities such as eye blinks have significant affect on the quality of the recorded signals (Figure \ref{fig:recording}), which makes intention recognition difficult.
\begin{figure}[htb]
\subfloat{
    \includegraphics[width=.275\textwidth]{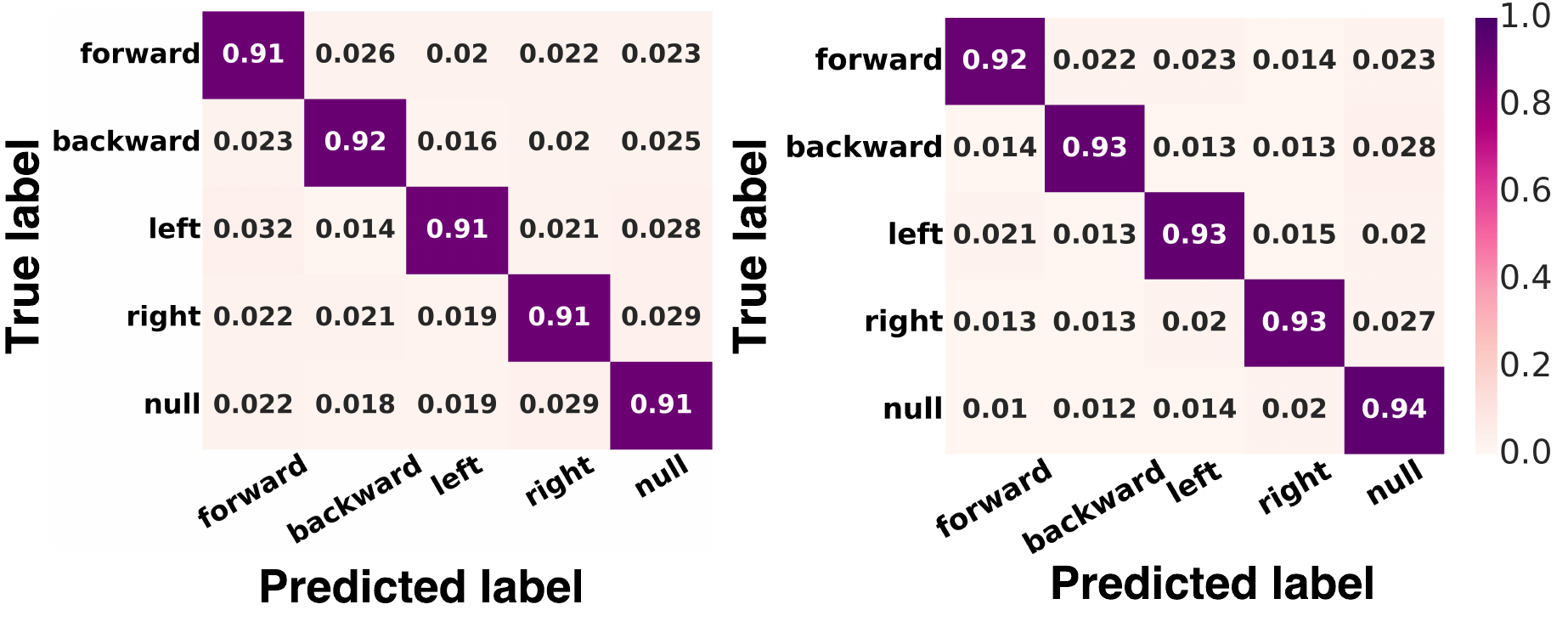}}
\subfloat{
    \includegraphics[width=.185\textwidth]{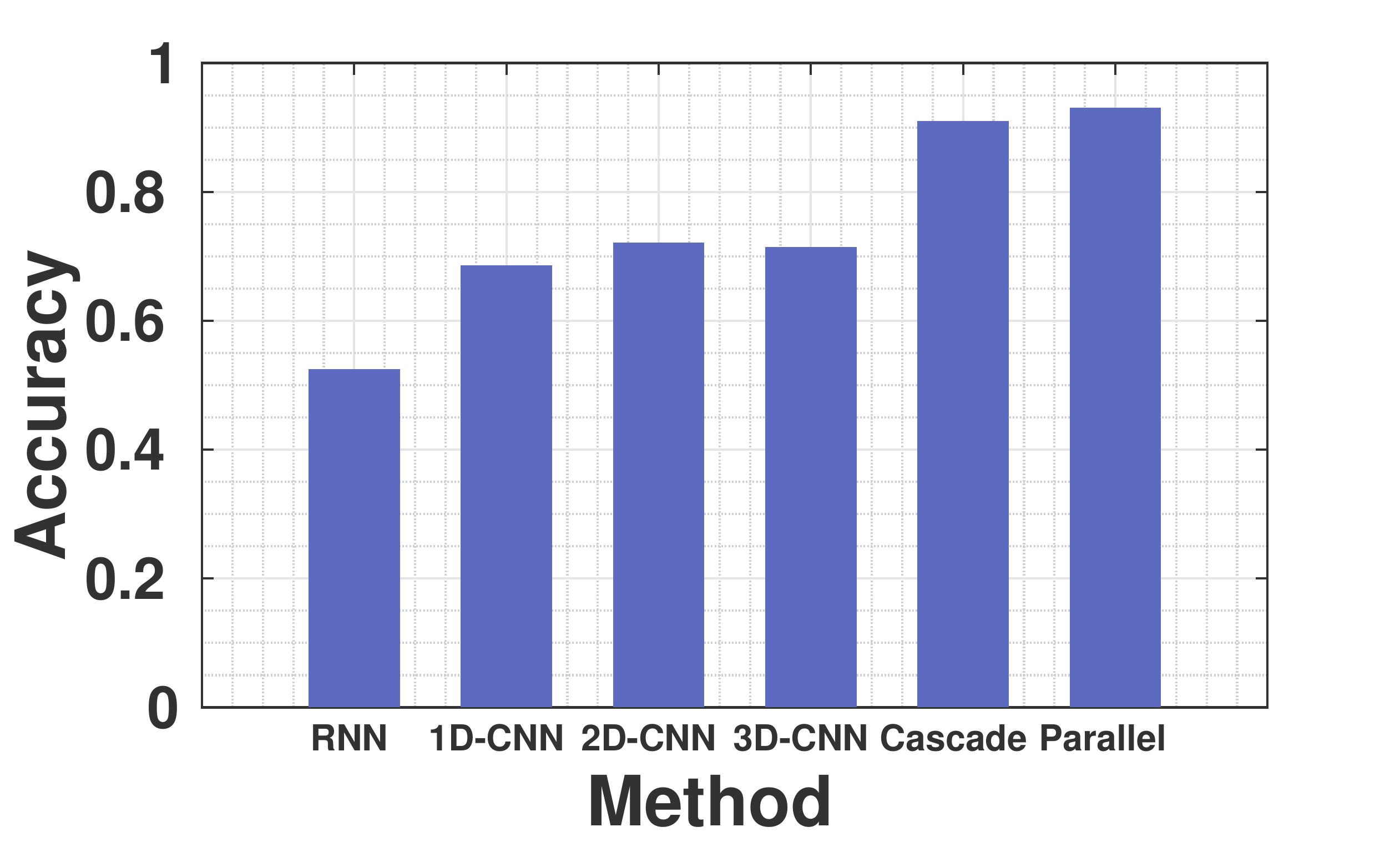}}
  \caption{Instruction intention recognition results (a) Confusion matrix of cascade model (b) Confusion matrix of parallel model (c) Performance comparison}
  \label{fig:casestudy}
\end{figure}
The instruction intention recognition is of more practical significance for general BCI applications. Figure \ref{fig:casestudy} shows the evaluating results on the case study dataset. Both the cascade and parallel models achieve excellent recognition accuracy higher than 90\%. The parallel model obtains the highest accuracy of 93.1\%, surpassing the best baseline model by more than 20\%. It is also noticed that the 2D-CNN model outperforms the 1D-CNN model, emphasizing the importance of spatial information for recognizing human intentions. Unexpectedly, the 3D-CNN model performs almost the same as the 2D-CNN model. This is probably due to the local temporal representations are of less effect on EEG signal analysis. However the global temporal information induced by the cascade and parallel models enhance the recognition performance considerably. We notice that the resulting performance is marginally lower than that using the PhysioNet dataset. This is due to the limited recording resolution of 14 EEG channels in our case study experiments compared with 64 recording channels in the PhysioNet dataset.  

\subsubsection{Demonstration}
The proposed framework was finally deployed on a customized BCI typing system. The alphabet is divided into clusters and instruction intentions of different directions are used to select different clusters. When one cluster is selected, its contained letters will be further divided until there is only one letter left in one cluster. \footnote{https://youtu.be/A9oqzNXejkg}

\section{Conclusions}
In this paper, we propose the use of spatio-temporal representations to enhance EEG-based intention recognition in a more practical scenario of cross-subject and multi-class, and develop two unified end-to-end trainable deep learning models for both movement intention and instruction intention recognition. 
Experiments on both the public dataset and the real-world BCI dataset demonstrate the effectiveness and feasibility of our models over diverse human intentions and various EEG resolutions. The variants of the proposed models and the influence of the spatio-temporal information are also systematically investigated. This work makes an important developing step toward accurate human intention recognition for practical BCI system research.

\bibliography{Bibliography-File}
\bibliographystyle{aaai}
\end{document}